\documentclass[12pt]{article}
\usepackage{cite}
\usepackage{epsfig}

\newcommand{\equ}[1]{Eq.\,(\ref{#1})}
\newcommand{\eqs}[1]{Eqs.\,(\ref{#1})}

\newcommand{\ew}{electroweak~}
\newcommand{\non}{\nonumber}
\newcommand{\gsim}{\;\rlap{\lower 3.5 pt \hbox{$\mathchar \sim$}} \raise 1pt
 \hbox {$>$}\;}
\newcommand{\lsim}{\;\rlap{\lower 3.5 pt \hbox{$\mathchar \sim$}} \raise 1pt
 \hbox {$<$}\;}

\newcommand{\msbar}{\overline{\rm MS}}

\newcommand{\smallz}{{\scriptscriptstyle Z}} 
\newcommand{\smallw}{{\scriptscriptstyle W}} %
\newcommand{\smallh}{{\scriptscriptstyle H}} %

\newcommand{\mz}{M_\smallz}
\newcommand{\mw}{M_\smallw}

\newcommand{\mh}{M_\smallh}
\newcommand{\mt}{M_t}
\newcommand{\mtbar}{\overline{M}_t}

\def\pl#1#2#3{{\it Phys. Lett. }{\bf B#1~}(19#2)~#3}
\def\zp#1#2#3{{\it Z. Phys. }{\bf C#1~}(19#2)~#3}
\def\prl#1#2#3{{\it Phys. Rev. Lett. }{\bf #1~}(19#2)~#3}

\def\pr#1#2#3{{\em Phys. Rev. }{\bf D#1~}(19#2)~#3}
\def\np#1#2#3{{\em Nucl. Phys. }{\bf B#1~}(19#2)~#3}

\newcommand{\be}{\begin{equation}}
\newcommand{\ee}{\end{equation}}
\newcommand{\een}{\end{subequations}}
\newcommand{\ben}{\begin{subequations}}
\newcommand{\beq}{\begin{eqalignno}}
\newcommand{\eeq}{\end{eqalignno}}
\newcommand{\bea}{\begin{eqnarray}}
\newcommand{\eea}{\end{eqnarray}}

\begin{document}
\boldmath
\unboldmath


\thispagestyle{empty}
\rightline{CERN-TH/2000-211}
\rightline{TUM-HEP-381/00}
\rightline{July 2000}
\vspace*{1.2truecm}
\bigskip

\centerline{\LARGE\bf  Electroweak  effects  in radiative B decays}
\vskip1truecm
\centerline{\large\bf Paolo Gambino$^a$ and Ulrich Haisch$^{b,c}$}
\bigskip
\begin{center}{
{\em $^a$ CERN, Theory Division, CH--1211 Geneve 23, Switzerland. }\\
\vspace{.3cm}
 {\em $^b$ Max-Planck-Institut f\"ur Physik
   (Werner-Heisenberg-Institut),\\
F\"ohringer Ring 6, 80805 M\"unchen, Germany}\\
\vspace{.3cm}
{\em $^c$ Technische Universit\"at M\"unchen, Physik Dept.,\\
James-Franck-Str., D-85748 Garching, Germany}
}\end{center}
\vspace{1.5cm}

\centerline{\bf Abstract}
\vspace{1.cm}
We compute  the  two--loop \ew corrections to 
the radiative decays of the $B$ meson in the SM.
Electroweak  effects
reduce  the Wilson coefficient $C_7^{eff}(\mw)$ by  $2.6$\%
for a light Higgs boson of about 100 GeV 
and are less important for a heavier Higgs.
The leading term of a heavy top expansion of our result differs
from the one obtained in the   {\it gaugeless}
approximation where only top quark Yukawa couplings are considered: 
we discuss the origin of the discrepancy and provide a 
criterion for the validity of the {\it gaugeless} approximation.
As a byproduct of the calculation we also obtain the $O(\alpha)$ 
corrections to the Wilson coefficient of the four--fermion operator $Q_2$.
A  careful analysis of the interplay between \ew and QCD effects
leads to  an overall 2\% reduction of the total branching ratio for
$B\to X_s \gamma$ due to purely \ew corrections.
For a light Higgs boson, the up--to--date
SM prediction  is $BR_\gamma=3.29\times 10^{-4}$. 
\vspace*{2.0cm}


\newpage

{\bf 1.} Radiative B decays represent one of the most important probes of new
physics and a major  testing ground for the Standard Model (SM). 
They already place severe constraints on many new physics
scenarios.
The present experimental accuracy for the branching ratio  of 
$B\to X_s \gamma$ (BR$_\gamma$ in the following) is
about 15\% \cite{cleo} and is expected to improve significantly
in the near future, both at CLEO and at the $B$ factories.

On the theoretical side, since precise predictions in the SM are
particularly important,  the subject has reached a high degree
of technical sophistication. Indeed, perturbative QCD corrections are very 
sizeable \cite{bert} and give the dominant contribution; 
they are best implemented in the framework of an effective
theory  obtained 
by integrating out the heavy degrees of freedom characterized by a
mass scale $M\ge \mw$. 
At lowest order in this approach the FCNC processes  $B\to X_s \gamma$
and $B\to X_s g$  proceed 
through helicity violating amplitudes induced by the magnetic operators 
\be
Q_{7}= \frac{e}{4\pi^2} m_b \,\bar{s}_L \sigma^{\mu\nu} b_R \, F_{\mu\nu}
\,,\ \ \ \ \ \ \
Q_8=\frac{g_s}{4\pi^2} m_b \,\bar{s}_L \sigma^{\mu\nu} t^a 
b_R \, G^a_{\mu\nu}\, .
\label{ops}
\ee
  A few years ago the
renormalization group improved QCD calculation has been completed at the
next-to-leading order (NLO) 
\cite{nlo1,nlo2,noi,nlo3}, reducing  the uncertainty 
from uncalculated QCD higher orders to about 5\%. More
recently, NLO predictions have been made available in some new physics
models as well \cite{noi,new}. 

There has also been progress  concerning
 QED and \ew radiative corrections:
after Czarnecki and Marciano considered all the leading QED logarithms
\cite{marciano},
their interplay with QCD corrections has been studied in
\cite{kagan,misiakQED}, and  Strumia \cite{strumia}
has calculated the leading
term of the Heavy Top Expansion (HTE) of the \ew two-loop corrections using the
 {\it gaugeless} limit of the SM.
As for non-perturbative effects, they seem to be under control \cite{nonpert},
although some aspects may still need  a more detailed investigation
\cite{misiaktalk}. 

In addition to uncalculated radiative corrections and non-perturbative
effects, the calculation of  BR$_\gamma$ 
 is also affected by  the uncertainties on the input
parameters (the CKM matrix elements, the semileptonic branching ratio
BR$_{SL}$,  etc.). In fact, the latter bring
the overall theoretical error to about 10\%.  As the parametric
uncertainties  (especially those on the CKM elements, $\alpha_s$,
BR$_{SL}$,  and $\mt$) are expected to  decrease soon, 
and given  the crucial importance of this decay mode, it seems appropriate 
to try and  refine the SM prediction as much as possible.

In this note we  reexamine the two-loop \ew contributions to
radiative $B$ decays and present the result of a  calculation
where only some photonic effects have been neglected. Moreover,
we  update the SM prediction of BR$_\gamma$ using the latest
experimental inputs. The final
result is expressed by  a compact formula that summarizes the
dependence on the input parameters.

{\bf 2.} Although generally small, two-loop purely 
\ew effects are sometimes very
important: an example is provided by the precision observables of  the
SM, like the effective sine measured on the $Z^0$ pole and the $W$ mass,   
where radiative corrections up to
$O(g^4 \mt^2/\mw^2)$ \cite{dgv,dg} are now routinely included in the
analysis with important  consequences in the \ew fits
\cite{dg,LEP}. Moreover, by  fixing the normalization of the \ew coupling,
two-loop  effects  reduce the \ew 
scheme dependence of the SM prediction, which can be quite large ---
also for FCNC processes \cite{bbbar}.  

As mentioned above, in the calculation of BR$_\gamma$ the leading
large logarithms of QED origin are now under
control, as a resummation of all $ \alpha \alpha_s^{n-1} (\ln
m_b/\mw)^n$ terms has been completed \cite{marciano,kagan,misiakQED}.
Apart from that, our knowledge of \ew effects in $b\to
s \gamma$  is limited to the  subset of two--loop fermion loop corrections
calculated in \cite{marciano} and to the leading term of the HTE
of \cite{strumia}. In fact, the two results are numerically very different
--- about $-2.3$\% and less than $-0.7$\%, respectively, on the Wilson
coefficient at $\mw$. 
The leading term of the HTE was calculated in 
 \cite{strumia} using the  {\it gaugeless} limit of the SM, i.e. in a 
Yukawa theory where the heavy top couples only to the Higgs doublet,
setting $\mw=0$ and keeping the Higgs mass $\mh$ finite and arbitrary. 
In the presence of external gauge bosons, these can be considered as
background sources. This approach presents a
few limitations that also motivate our new calculation: 

\begin{itemize}
\item the lowest order contribution to the Wilson coefficient
of  $Q_7$  is a function of the top mass
whose HTE converges very slowly. Using $x_t= \mt^2/\mw^2\approx 4.7$
and writing explicitly the numerical values of the successive
$O(1/x_t^n)$ terms, it reads
\bea
C_7^{(0)}(x_t)&=& \frac{x_t(7-5x_t -8 x_t^2)}{24(x_t -1)^3} + 
\frac{x_t^2(3x_t - 2)\,\ln x_t}{4 (x_t -1)^4}=\\
&=& -\frac13 -0.010 + 0.070 +0.046 +0.021 +...= -0.195\non
\eea
where the ellipses represent  contributions $O(1/x_t^5)$ or higher.
The leading HTE is  therefore unlikely to provide anything more 
than an order of magnitude estimate 
of the two-loop \ew contribution. In this respect, the 
similar case of $B_0 - \bar{B}_0$ mixing  \cite{bbbar} 
is very instructive: for
realistic values of the top mass the complete two-loop \ew correction
is not well  approximated even by the first three terms of the HTE and the
leading HTE term is numerically far from the complete result.
 
 \item even  assuming the leading HTE term 
to be representative, it should not  be expected to give  an accurate 
result for a light Higgs mass,  $\mh \approx O(\mw)$, because it is 
 obtained by setting  $\mw=0$ \cite{dg}. 
On the other hand, present \ew fits show a decisive preference for 
a light Higgs boson,  $\mh < 215$ GeV at 95\% C.L. \cite{LEP}.

\item   the {\it gaugeless} limit  has often been used 
to compute the leading HTE term, but it is known \cite{hzz}
that in some cases 
it does {\it not} reproduce the correct result. In the following
we explain why it fails   for
radiative $B$ decays and provide a general criterion for its use.

\end{itemize}

A complete calculation of all \ew effects in radiative 
$B$ decays in the framework of effective Hamiltonians
is a very complex enterprise which  involves other operators in addition
to those of \equ{ops}. In fact, the analysis should be aimed at resumming
all $\alpha\alpha_s^n \,(\ln m_b/\mw)^{n} $ effects. The procedure is
summarized, for instance, in \cite{penguins}.
Its necessary steps would be: 
(i) the calculation of two--loop $O(\alpha)$ 
matching  conditions for $Q_{7,8}$ 
at some $O(\mw)$ scale --- this involves also their QED mixing with all 
other operators --- and of the $O(\alpha)$ contributions to various 
four quark operators;
(ii) QED--QCD running of the Wilson coefficients to the $B$ mass scale
--- this would require a three loop computation of the anomalous
dimension matrix similar to that of \cite{nlo1};
(iii) calculation of the one-loop QED matrix elements of the various
operators --- the  determination of these matrix elements depends
sensitively on the precise experimental conditions.

An important simplification can be obtained by keeping only the 
first term in an expansion around $s_\smallw=\sin\theta_\smallw=0$. This is
equivalent to considering a $SU(2)_L$ theory with a background photon field 
and removes all the light virtual degrees of freedom.
In particular, all 
the diagrams with virtual photons and all infrared
(IR) divergences drop out of the two--loop calculation in a gauge--invariant
way. Step (i) is therefore 
much simpler as the calculation of the two-loop $b\to s \gamma $ and 
$b\to s g$ amplitudes  gives us directly the scheme independent 
$O(g^2)$ correction to
$C_{7,8}$, respectively ($g=e/s_\smallw$ is the $SU(2)_L$ coupling). 
Moreover, this simplification avoids completely
steps (ii) and (iii), because they are both driven by purely photonic effects
suppressed at least by $Q_u |Q_d|  s_\smallw^2\approx 0.05 $ 
with respect to pure $SU(2)_L$ contributions.

In analogy to \cite{penguins}, we complement this 
approximation scheme by keeping also the $O(g^2 \mt^2/\mw^2)$
contributions that vanish  as $s_\smallw\to 0$.
In practice, we therefore expand the two--loop $b\to s \gamma (g)$
amplitude $A^{(2)}$ in powers of $s_\smallw^2$:
\be
A^{(2)}= g^4 \left[ A_0 + A_1 s_\smallw^2 + O(s_\smallw^4)\right]
\ee
and retain only $A_0$ and the $O(\mt^2)$ part of $A_1$.
This is likely to be a sufficiently good approximation, as
suggested by  those  cases 
\cite{bbbar,dffgv} where  it has been possible to compare it with  the 
complete SM result. We recall that $\Delta\rho$ \cite{dffgv} 
and $B_0-\bar{B}_0$ mixing \cite{bbbar} involve amplitudes
conceptually  similar to those under consideration: 
forbidden at tree level and induced at one--loop by virtual $SU(2)_L$ effects.
Later on, we will give an estimate of the
residual uncertainty.

We calculate analytically the two-loop amplitudes in the Feynman
background gauge 
with a background photon (gluon). 
A few thousand diagrams  are automatically generated
by the package {\it FeynArts 2.2} \cite{feynarts} (the topologies are
shown in Fig.~1).
\begin{figure}[h]\vspace{-2.7cm}
\begin{center}
\mbox{\epsfxsize=10cm\epsffile{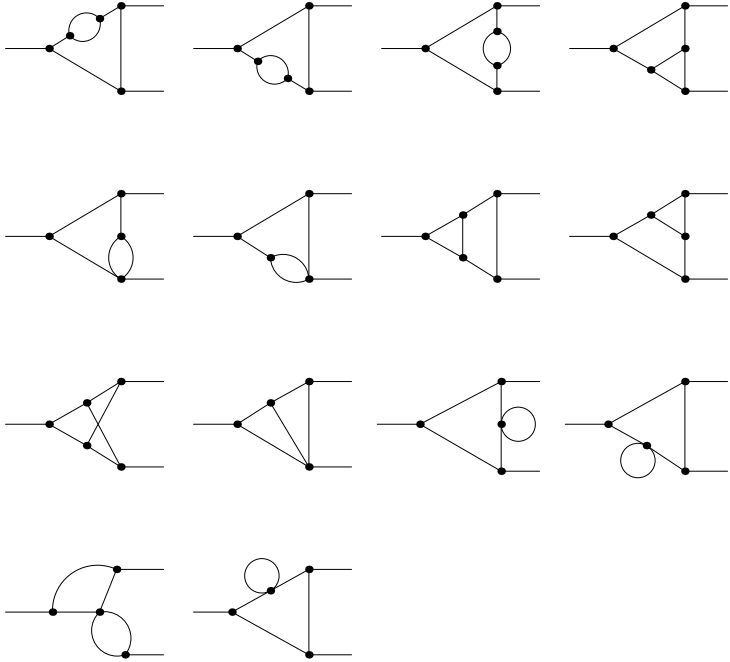}}
\end{center}
\vspace{-2cm}
\caption{\sf Two-loop topologies for $b\to s \gamma$.\label{fig1}}
\end{figure}
After setting to zero all light fermion masses but the 
$b$ quark mass, they  can be reduced to a few hundred
equivalence classes, which we have actually computed. Due to 
the GIM mechanism, the  CKM cofactor of each
equivalence class is always proportional to $\lambda_t= V_{tb} V_{ts}^*$.
The extraction of the magnetic penguin amplitude and the two-loop integration
are  performed as in \cite{noi}.
All the steps of the calculation have been implemented in two 
independent and completely automatic codes that involve various
combinations of {\sc Mathemathica} \cite{math} and {\sc Form}
\cite{form} routines. Although the result can be expressed in terms 
of logarithms and dilogarithms, it is rather lengthy and we will
present instead accurate numerical approximations.

A peculiarity of  the two-loop calculation for these  processes is
 the presence of diagrams containing anomalous
fermionic loops (triangles). It is well  known that the naive
definition of anticommuting $\gamma_5$ in $n$ dimensions that we
employ in the rest of the calculation fails for these diagrams
because it  leads to algebraical
ambiguities and cannot reproduce the axial anomaly. Our solution consists
in calculating Dirac structures containing an odd number of
$\gamma_5$'s --- i.e. those leading to the anomalous term ---
 using  anticommuting $\gamma_5$ in strictly four dimensions, 
which is possible because of
their apparent  UV convergence. The anomaly cancellation then
guarantees the absence of both anomalous and ambiguous terms 
in the sum of all diagrams (see \cite{martin}, app. C).
We have also checked our results for these specific terms
using the HV definition of $\gamma_5$ in $n$ dimensions \cite{HV}. 
From a formal point of view, the
equivalence of the two methods follows from the absence of
non-invariant counterterms for the odd $\gamma_5$ part in
 the HV case \cite{martin}, as can be
also  seen in full generality using the powerful formalism  of \cite{antonio}.

The renormalization is performed following the simple framework of
 \cite{bbbar}, to which we refer for a detailed discussion
(see also \cite{dg}) and for the notation. 
We recall that the top mass is renormalized on-shell as
 far as \ew effects are concerned, but it is customary to use an $\msbar$
definition for the QCD effects. Although the choice of the scale
$\mu_t$ for the $\msbar$ top mass is a matter of convention, the NLO QCD
corrections depend sensitively on $\mu_t$. For simplicity, we follow 
\cite{noi} and in the numerics we set $\mu_t=\mu_\smallw=\mw$ and  employ
$\mtbar\equiv\mtbar(\mw)=175.5\pm 5.1$ GeV obtained from the pole mass value
$\mt^{pole}=174.3\pm 5.1$ GeV \cite{pdg}. No renormalization of the electric charge 
is necessary in our approximation. The renormalization of the $b$ quark mass,
not needed in \cite{bbbar}, is also performed on-shell. The
 counterterm reads
\bea\non
\frac{\delta m_b}{m_b}= 
{g^2 \over 16 \pi^2}  \left ({\bar \mu^2 \over \mtbar^2}
\right ) ^\epsilon \left[ \frac3{8\epsilon} \left(x_t-1 \right) 
-   \frac{3 + 8x_t - 5x_t^2}{16\,\left( x_t-1 \right) } - 
   \frac{3\,\left( 1 - 3x_t + x_t^2 \right) \ln x_t}
    {8\,{\left( x_t -1\right) }^2}
 + O(s_\smallw^2)
\right]\non .
\eea
One should keep in mind that 
 the $m_b$ factors in \equ{ops}  originate either from the $b$-quark
Yukawa coupling or from  the use of on-shell equations of motion.
In the latter case, $m_b$  should {\it not} be
renormalized as it is on-shell by definition. This is the mass appearing
explicitly in the projector of Eq.(14) of \cite{noi}.  
 Indeed, besides the magnetic operators of \equ{ops},
there are additional off-shell operators that project onto $Q_{7,8}$
when the external momenta are set on-shell, i.e.\, 
$Q_{10} \sim e\, {\bar s}_L 
\{ \not \!\!D, \sigma^{\mu\nu} \}b_L \, F_{\mu\nu}$ and the
analogous one with gluon fields \cite{noi}. 
In correcting the external fields, one should take into account 
that the chirality of the $b$ quark is different in $Q_{7}$ and
$Q_{10}$. The external leg corrections 
of \cite{bbbar} correspond to the correct LSZ factors and 
implement the renormalization of the CKM matrix according to
\cite{CKM} within our approximations. We recall that this  gauge invariant
definition of the CKM matrix is the most appropriate to the present 
low-energy measurements because, unlike an $\msbar$ renormalization,
 it avoids $O(g^2)$ corrections not suppressed by GIM and 
proportional to $(m_i^2+m_j^2)/(m_i^2-m_j^2)$, where $m_{i,j}$ are
light quark masses \cite{CKM}.   
Notice also that in our framework $\delta (Z^R_d)_{ij} =0$. 

According to the standard procedure, we will normalize BR$_\gamma$
  to the semileptonic branching ratio, BR$_{SL}$. 
This fixes the normalization of the
\ew coupling but requires the inclusion of the
one-loop \ew corrections to BR$_{SL}$ \cite{si78}. It is
straightforward to see that, up to $O(s_\smallw^2)$ terms that we
neglect, these are the same that enter the muon decay. Hence, in this
  respect the
use of BR$_{SL}$ is effectively equivalent to that of the Fermi constant
measured in muon decays, $G_\mu$, and the coupling 
renormalization proceeds as described for this case 
in \cite{bbbar}. We incorporate the complete one-loop 
correction to the muon decay amplitude, without taking the 
$s_\smallw \to 0$ limit, on the ground that this is an independent 
process for which the complete correction is available. Notice that the 
leading part of the photonic corrections to BR$_{SL}$, characterized
  by large logarithms and  not considered in our calculation 
is part of the $O(\alpha\alpha_s^{n-1} \ln^n m_b/\mw)$ 
analysis of \cite{marciano,kagan,misiakQED} and
is included in our numerical results.

We now recall that the regularization scheme--independent quantity entering the
calculation of BR$_\gamma$ is not $C_7(\mu_b)$ but a combination 
$C_7^{eff}(\mu_b)$
of this Wilson coefficient and of the coefficients of the four fermion
operators with mixed chirality \cite{lectures,nlo1}. The case of QED
effects has been also considered in  \cite{misiakQED}. 
It turns out that,
 as far as \ew corrections are concerned, the two scheme--independent
 quantities relevant for $B\to X_s \gamma $ and $B\to X_s g$ are
\be
C_7^{eff}(\mu)= C_7(\mu) +\frac16 C_7^{ P}(\mu) +\frac12 C_8^{ P}(\mu),
\ \ \ \ \ 
C_8^{eff}(\mu)= C_8(\mu) -\frac12 C_7^{ P}(\mu),
\label{cieff}
\ee
where $C_{7,8}^{ P}$ are the  Wilson coefficient of two \ew penguin 
operators 
\bea
Q_7^P= {3\over 2}\;(\bar s d)_{V-A} \! \! \sum_{q=u,d,s,c,b} \! \! e_q\;(\bar
qq)_{V+A} \, ,
\ \ \ \ \
Q_8^P={3\over2}\;(\bar s_{\alpha} d_{\beta})_{V-A} \! \!
\sum_{q=u,d,s,c,b} \! \! e_q
        (\bar q_{\beta} q_{\alpha})_{V+A}.
\label{penguin}
\eea
Notice that in QCD  there are other contributions to
\equ{cieff}  \cite{nlo1}.
As $C_8^{P}(\mw)=0$ at leading order and  $C_7^{P}(\mw)$ is proportional to
$s_\smallw^2$, we need to  consider only the part of 
$C_7^{P}(\mw)$ enhanced by $\mt^2$ (which actually approximates the
full Wilson coefficient very well \cite{penguins}).

Our  results for the $O(g^4)$ contributions to $B\to X_s \gamma (g)$ 
can therefore be written
 as  additional contributions to $C_{7,8}^{eff}(\mw)$ and are accurately
approximated by 
\bea
\delta C_{7,ew}^{eff}= \frac{g^2}{16\pi^2} \left[
1.8615 - 2.422 \left(1-\frac{\mtbar}{175.5}\right) - 0.4463 
  \ln\frac{\mh}{100} - 0.216 \ln^2\frac{\mh}{100}
 \right]&&\non\\
\delta C_{8,ew}^{eff}= \frac{g^2}{16\pi^2} \left[ 
0.2596  +0.282 \left(1-\frac{\mtbar}{175.5}\right)
             - 0.1366  \ln\frac{\mh}{100} - 0.021 
\ln^2\frac{\mh}{100}
\right]\label{fits}&&
\eea
where the $SU(2)_L$ coupling $g$ can be calculated 
 from the relation $g^2=4\sqrt{2} G_\mu \mw^2$.
We use $\mw=80.419$ GeV and $s^2_\smallw=0.23145$ for the $O(g^2
 s^2_\smallw \mt^2)$ contributions that we retain.
 \eqs{fits} reproduce accurately (within 
1\%) the analytic results 
in the ranges $100<\mh<250$ GeV and $165<\mtbar<180$ GeV.
We stress that \eqs{fits} are independent of the choice of the scale
 $\mu_t$ in
 the QCD top mass definition: it is sufficient to calculate
 $\mtbar(\mu_t)$ and employ it in \eqs{fits}. Differences between
 different choices are present in the QCD corrections to BR$_\gamma$
but are  higher order effects as far as the present calculation is concerned.

The numerical relevance of our corrections to $C^{eff}_{7,8}(\mw)$
is shown in Fig.\,2 for  $\mtbar=175.5$ GeV: at $\mh=100$ GeV the Wilson
 coefficients of $Q_{7,8}$ are reduced, respectively by 
2.6\% and 0.7\%.
\begin{figure}[t]
\begin{center}
\begin{tabular}{cc}
\hspace{-0.5cm}\mbox{\epsfxsize=8cm\epsffile{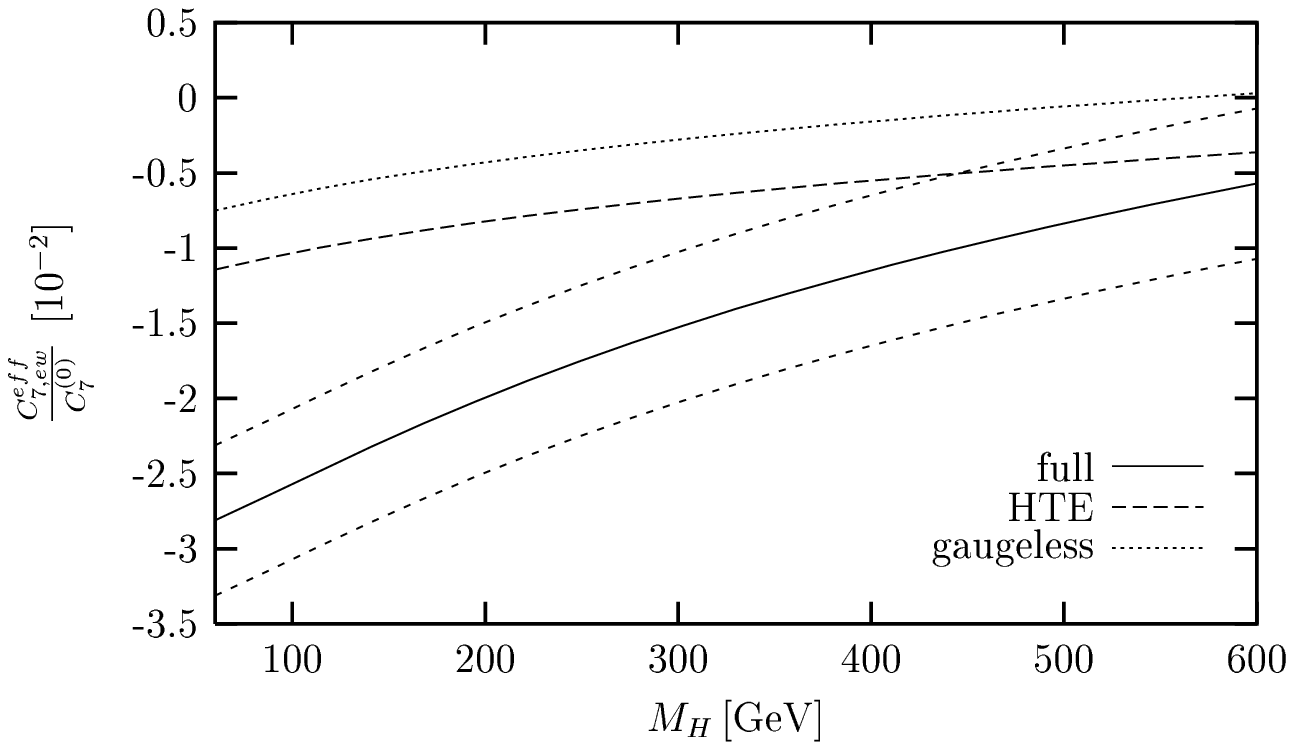}}&
\hspace{-0.5cm}\mbox{\epsfxsize=8cm\epsffile{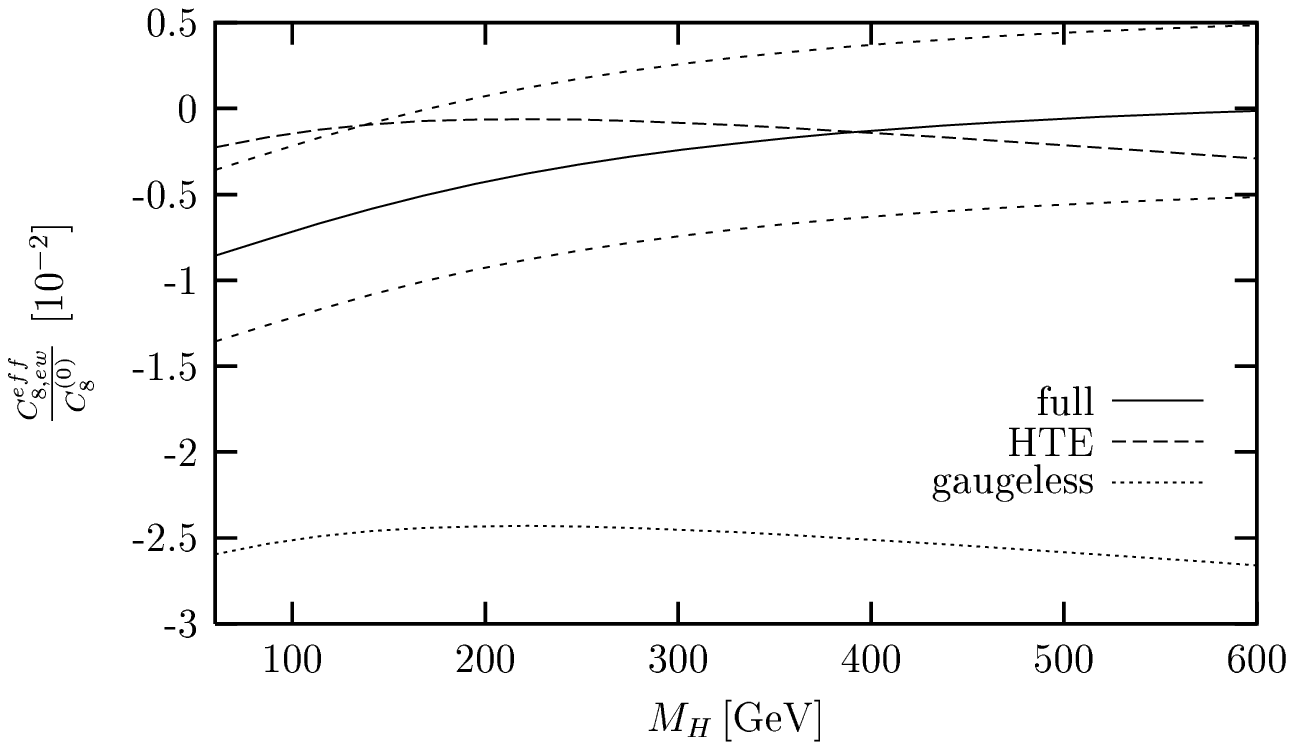}}
\end{tabular}
\end{center}
\vspace{-0.5cm}
\caption{\sf Electroweak corrections to the Wilson coefficients
$C_{7,8}(\mw)$. The solid lines represent our results with their 
error estimates, the dashed lines their leading HTE, and the dotted lines
the results of the {\it gaugeless} approximation.\label{fig2}}
\end{figure}
As a measure of the uncertainty due to the expansion around
 $s_\smallw=0$ we use the difference between the complete
correction to the muon decay and its $s_\smallw\to 0$ limit, which
amounts to about 0.5\%. This seems to us a realistic
estimate of the error due to our approximation and we will use it in 
the following.
If we consider only  fermionic loops we reproduce the results of
\cite{marciano} for $C_7$, which lead to a $-2.3$\% reduction of
$C_7(\mw)$. Although purely accidental, the closeness of this fermion
loop approximation to our complete result for a light Higgs,
$\mh\approx 100$ GeV is impressive.

{\bf 3.} Let us now consider the HTE of our results and see how it
compares with existing analyses. In units 
$g^2/(16\pi^2) \ \mtbar^2/(2\mw^2)$ it is given by
\bea
\delta C_{7,{\rm HTE}}^{eff,ew}=
\frac{55h_t -16 - 11{h_t^2} - 26{h_t^3}}{144h_t} - 
\frac{2h_t-16 - 36{h_t^2} + 74{h_t^4} - 45{h_t^5} + 2{h_t^6} }{864
  h_t^2} \,\pi^2 +\frac{s^2_\smallw}{27}  \non\\
 -\,\frac{8 - h_t - 6{h_t^2} - 52{h_t^3} + 85{h_t^4} - 33{h_t^5} +
  2{h_t^6} }{72{h_t^2}}\, {\rm Li_2}(1 - h_t) 
-\frac{74 - 45h_t + 2 h_t^2}{288} h_t^2 \ln^2 h_t \ 
\non\\ 
 - \,\frac{80 + 68h_t - 262{h_t^2} + 134{h_t^3} - 25{h_t^4} + 2{h_t^5}
  }{288h_t}
\,\phi \left(\frac{h_t}{4}\right)
+\frac{8  - 17h_t - 2 h_t^2 - 14{h_t^3} }{72h_t}\ln h_t \ \ 
 \label{HTE}
\eea
and 
\bea
\delta C_{8,{\rm HTE}}^{eff,ew}=
  \frac{32 - 83\,h_t - 23\,h_t^2 + 
      16\,h_t^3}{96\,h_t} - 
   \frac{ 8 - h_t - 18\,h_t^2 - 
        h_t^4 + 9\,h_t^5 - h_t^6
         }{144\,h_t^2}\,{\pi }^2 - \frac{ s^2_\smallw}{9} \ \ \ \ \ 
       \ \non\\
+  \,  \frac{16 - 2h_t - 12h_t^2 + 
        40h_t^3 - h_t^4 - 
        30h_t^5 + 4h_t^6 }{48\,h_t^2} {\rm Li}_2(1 - h_t)+
 \frac{
      1 - 9 h_t + h_t^2  }{48}h_t^2\ln^2 h_t \non\\
+ \, \frac{8 - 58\,h_t + 62\,h_t^2 + 
        17\,h_t^3 - 16\,h_t^4 + 
        2\,h_t^5 }{96\,
      h_t} \,\phi\left(\frac{h_t}{4}\right) - 
   \frac{8 + h_t + 7h_t^2 - 5h_t^3 }{24\,
      h_t}\,\ln h_t\ \ \ \,    
\label{HTEc8}
\eea
where $h_t=\mh^2/\mtbar^2$,  ${\rm Li_2}(x)$ is the
dilogarithmic function, and $\phi(x)$ is given in Eq.\,(48) of \cite{bbbar}.
As can be seen in Fig.\,2, the leading HTE term
approximates our full result very poorly, especially for a light Higgs. 
We have also studied the convergence of the HTE, calculating its 
first three terms, and found that for realistic $\mt$ values they do not
converge, in a way very similar to \cite{bbbar}.  
Our \equ{HTE} differs from the analogous one in \cite{strumia} by a term 
$\frac29\left(\frac12 -\frac{ s_\smallw^2}{3}\right) +\frac{s_\smallw^2}{9} $. 
On the other hand, we agree with
\cite{strumia} if we perform the calculation in the {\it gaugeless}
limit. This is not surprising because it is known \cite{hzz} that 
the {\it gaugeless} approximation does not always include all  leading 
$\mt^2$ contributions. 

To understand better this point, notice that for a asymptotically 
heavy top  both the top Yukawa coupling, $g_t=g\, 
\mt/(2\mw)$,  and 
the loop integration can provide powers of the top mass. 
In the case at hand, the one--loop integrals are
convergent, so that the one--loop contributions scale at most like 
$g_t^2/\mt^2\sim g^2/\mw^2$. At the two--loop level, the {\it
  gaugeless} contributions scale as
$g_t^4/\mt^2\sim g_t^2 \ g^2 /\mw^2\sim g^4 \mt^2/\mw^2$, but the same 
heavy top behaviour can be obtained by inserting a dimension four
operator\footnote{Dimension two insertions are removed by our choice
  of renormalization in the $W-\phi$ sector.}  
proportional to $g_t^2$ in a topless loop. 
In general, the effective lagrangian
obtained after integrating out the heavy top  tells us exactly which
 the relevant  operators  are \cite{fer}. 
In our case, only the diagrams in Fig.\,3 contribute to the leading HTE
through the insertion of a flavor changing dimension four $Z^0$
penguin operator of the kind $\bar{s}_L \gamma^\mu  b_L Z^0_\mu$. 
The diagram with a mass insertion on the internal $b$
line depends on the regularization scheme --
it vanishes if IR divergences are regulated dimensionally -- and in the schemes
where it does not vanish it 
is cancelled in the matching by a contribution from the \ew penguin
operator $Q_7^P$ of \equ{penguin}. In both  cases, however, 
its contribution is  reintroduced in the 
quantity $C_{7,8}^{eff}$ by $C_7^P$.
In the limit of a heavy top the 
effective vertex has the  form
\be
\Gamma_{\bar{s} b Z^0}^\mu = i \frac{g^3}{(16 \pi^2)} 
\frac{\lambda_t}{c_\smallw}    
        \,\frac{x_t}{4} \, \bar{s}_L \gamma^\mu  b_L    
\ee  
with  $c_\smallw=\cos\theta_\smallw$. 
Inserting this gauge-independent 
effective coupling in the one-loop diagrams of Fig.\,3, and
keeping in mind  the tree level 
couplings of the $Z^0$ boson with $b_L$, $\sim (1/2 -
s_\smallw^2/3)$, and  with $b_R$, $\sim
-s_\smallw^2/3$,  we obtain  the
difference between the HTE of our result and the {\it gaugeless} limit.
\begin{figure}[t]
\begin{center}
\mbox{\epsfxsize=13cm\epsffile{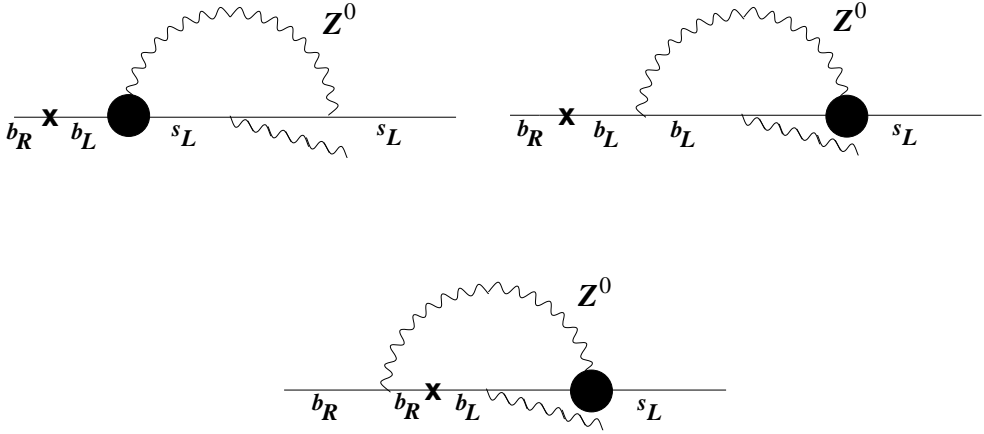}}
\end{center}
\vspace{-0.5cm}
\caption{\sf Insertions of an effective $Z^0$ penguin vertex in a
one-loop diagram.\label{fig3}}
\end{figure}
The argument is completely analogous for $C_8$, whose HTE also differs
from the {\it gaugeless} approximation.

So when does  the {\it gaugeless} limit potentially fail at two-loop?  
Whenever at the one-loop level the top quark diagrams in the limit of 
a heavy top scale like  a constant, namely in the same way as
the topless contributions. Indeed, in
this case we know that there are some dimension four effective operators
proportional to $g_t^2$ that can be inserted in  
one-loop diagrams not containing the top
and give contributions of the same order, in the limit of  heavy
top, of those belonging to the two-loop {\it gaugeless} approximation.

Table 1 summarizes the situation for the processes
considered in the literature at the two-loop level. 
It should be clear that the {\it gaugeless}
limit works safely only when the asymptotic expansion in  $\mt$ has maximal 
power ($\mt^2 $ at one-loop, $\mt^4$ at two-loop). 
Of course, there might be exceptions.
Indeed, whether the $O(g_t^2)$ dimension four operators are relevant or not
depends on the process under consideration. For instance, 
in the $Hgg$ effective vertex \cite{hgg} --- relevant for gluon--gluon
fusion production
and hadronic decays of the SM Higgs boson --- they are not because the 
gluons have no \ew interaction. This is in contrast to the similar
case of the 
$H\gamma\gamma$ effective vertex \cite{hzz}, where the {\it gaugeless}
approximation does not give the correct result. 
Similar considerations apply to the heavy Higgs limit, although the
leading term in the heavy Higgs expansion, subject  to other 
constraints, is not always what  is expected  from dimensional analysis.
\renewcommand{\arraystretch}{1.4}
\begin{table}[t]
\begin{center}
\begin{tabular}{|c|c|c|c|c|} 
\hline 
  & $b s \gamma,bsg$ &$H\gamma\gamma ,\,Hgg$  &
  $\Delta\rho,R_b,HZZ,K\!\!\to\!\pi\nu\bar{\nu}$
   &  $B_0$-$\bar{B}_0$ \\ 
 &  & \cite{hzz,hgg} & \cite{barb,hzz,buchalla:98} & \cite{bbbar} 
 \\ \hline
One loop            &
$\frac{g_t^2}{\mt^2}\sim \frac{g^2}{\mw^2}$ & 
$\frac{ (e^2\!,g_s^2)\, v \,g_t \mt}{\mt^2}\sim (e^2\!,g_s^2)$ & 
 $\frac{g^2\mt^2}{\mw^2}\,\,{\rm or}\,\, {g_t^2}$ & 
$\frac{g_t^4}{\mt^2}$    \\ \hline
Two loop           & 
$\frac{g_t^4}{\mt^2}\sim \frac{g^2g_t^2}{\mw^2}$ &
$\frac{(e^2\!,g_s^2)\, v \,g_t^3 \mt}{\mt^2}\sim (e^2\!,g_s^2)\, g_t^2$ & 
 $\frac{g^2 g_t^2\mt^2}{\mw^2}\,\,{\rm or}\,\, {g_t^4}$   & 
$\frac{g_t^6}{\mt^2}$    \\ \hline
\end{tabular} 
\caption{\sf Leading HTE contributions to different processes.
  Following  the
  criterion given in the text, the {\it gaugeless} limit fails in the
  first two cases.}
\end{center} 
\end{table} 

{\bf 4.} Let us  now examine in some detail 
the effect of our calculation on  BR$_\gamma$. 
As a first step we calculate the Wilson coefficients at a scale $\mu_b=4.8 $
GeV. It is well-known that the large mixing between $Q_2=
(\bar{c}b)_{V-A} \,(\bar{c}s)_{V-A}$ and $Q_7$
induces additive terms in the running of the coefficients from the $W$
to the $b$ mass scale which are numerically
very important.  Our aim is to resum all contributions 
$O(g^2 \alpha_s^n L^n)$ and
$O(\alpha \mt^2/\mw^2 \alpha_s^n L^n)$, where $L$ is a  large
logarithm. As mentioned above, these terms are uniquely originated by  heavy
degrees of freedom and enter only the determination of the Wilson
coefficients at a high scale $\mu\approx \mw$. At this order the
evolution of the coefficients is therefore driven only by LO QCD effects.  
The Wilson coefficient at the bottom mass scale is
given at LO in QCD by 
\be
C_7^{(0)eff}(\mu_b)= \eta^{\frac{16}{23}} C_7^{(0)eff}(\mw)
+\frac83 \left( \eta^{\frac{14}{23}}-\eta^{\frac{16}{23}}\right)
C_8^{(0)eff}(\mw) + C_2^{(0)}(\mw) \sum_{i=1}^8 h_i \eta^{a_i}
\label{running}
\ee
where $\eta= \alpha_s(\mu_\smallw)/\alpha_s(\mu_b)\approx 0.56$ and
$h_i, a_i$ are constants given e.g.\ in \cite{lectures}. The last term
is approximately equal to $-0.173$ and is dominant. 

The \ew corrections affect  \equ{running} in two ways: (i) 
they shift $C_i^{(0)eff}(\mw)$
by $\delta C_{i,ew}^{eff}$; (ii) they  introduce in \equ{running}
those gluon and \ew penguin operators that have non--zero $O(g^2)$
or $O(\alpha \mt^2)$ contributions to the 
Wilson coefficients at $\mu=\mw$. In the basis of \cite{lectures}, these are
$Q_{3,7,9}^P$. Their LO QCD mixing with the magnetic penguin operators 
can be gleaned from the anomalous dimension matrix
$\hat{\gamma}_s^{(0)eff}$ given in
\cite{roma,misiakQED} after a change of basis. The new entries of 
$\hat{\gamma}_s^{(0)eff}$ calculated in \cite{misiakQED} are given in
Table 2 in the conventional basis adopted in  \cite{lectures}.
The additional contributions to \equ{running} are therefore
\be
\delta^{ew} C_7^{(0)eff}(\mu_b)=
\eta^{\frac{16}{23}} \delta C_{7,ew}^{eff} +
\frac83 \left( \eta^{\frac{14}{23}}-\eta^{\frac{16}{23}}\right)
\delta C_{8,ew}^{eff} + 
\sum_{i=3,7,9} C^{P}_i(\mw) \sum_j h_{ij} \eta^{a_{ij}}
\label{ew}
\ee
where $C^{P}_i(\mw)$ are the relevant $O(\alpha)$ contributions
to the  Wilson coefficients and $h_{ij}, a_{ij}$ are magic numbers 
that can be easily determined from the anomalous dimension matrix.
The last term in \equ{ew} is approximately given by
$ 0.15 \,C_3(\mw) + 0.12 \,C_7(\mw) - 0.03 \,C_9(\mw)$
and is numerically very small; it  reduces
$C_7^{(0)eff}(\mu_b)$ by $-0.2$\%.
\begin{table}[t]
\begin{center}
\begin{tabular}{|c|c|c|c|c|} 
\hline 
$i$ &$ P7$ &$ P8$ &$ P9$ &$ P10$\\ \hline
$\hat{\gamma}^{(0)eff}_{s,i7}$& $-\frac{16}{9}$ &$-\frac{1196}{81}$ &
$\frac{232}{81}$ & $\frac{1180}{81}$\\ \hline
$  \hat{\gamma}^{(0)eff}_{s,i8}$& $\frac56$ &$ -\frac{11}{54}$ & 
$-\frac{59}{54}$ &$-\frac{46}{27}$ \\ \hline
\end{tabular} 
\end{center}
\caption{\sf Anomalous dimension matrix entries 
relevant for the mixing between \ew and magnetic penguins
\cite{misiakQED} 
in the basis of \cite{lectures}. Two $up$ and three $down$ active
flavors are assumed.}
\end{table} 

Notice now that  $C_2^{(0)}(\mw)$ in \equ{running} is unaffected by
\ew corrections of the kind considered here if $G_\mu$ is used to
normalize the effective Hamiltonian (as in fact we do); 
in that case $C_2^{(0)}(\mw)$ does
however  receive $O(g^2
s_\smallw^2)$ corrections.  In the NDR scheme we find 
\be
C_2(\mw)= 1+ \frac{{\alpha}(\mw)}{4\pi} \
 \left[ - \frac{22}{9} 
+\frac43  \ln \frac{\mz^2}{\mw^2} \right] + O(\alpha_s),
\label{c2}
\ee
where $\alpha(\mw)$ is the electromagnetic running coupling evaluated 
at $\mw$.
The $O({\alpha})$ \ew corrections to the Wilson coefficient $C_2$  are 
therefore very small
($-0.13$\%). But we stress that  a different choice of
normalization would induce additional (much larger) \ew contributions,
as can be easily seen using \cite{bbbar}. 
\equ{c2} is a new result that improves on \cite{c2}, where only QED
effects were taken into account, and includes all one-loop \ew
contributions.

From \equ{running} and neglecting the $O({\alpha})$ effects of \equ{c2}
we see that 
 $C_7^{(0)eff}(\mu_b)$ is reduced by only $(1.3\pm 0.2)$\%
for $\mh=100$ GeV due to the \ew corrections we have calculated; 
the reduction is  less pronounced for larger Higgs masses.
In a similar way, we find that for $C_8^{(0)eff}(\mu_b)$ the numerical
impact of the last term in \equ{ew} is more important than that of the two-loop
correction; \ew effects  increase  $C_8^{(0)eff}(\mu_b)$ 
by about $0.3\pm 0.2$\% for $\mh=100$ GeV.
In addition to the leading logarithmic QCD effects 
considered in \equ{running} there are
next-to-leading QCD terms that enter the calculation of BR$_\gamma$
\cite{nlo1}.  Our \ew corrections
affect them only as a NNLO effect. 
As there are other uncalculated contributions
to that order, even in the approximation we have adopted --- 
in particular  $O(g^2 \alpha_s)$ corrections to the Wilson
coefficients, as can be seen from  Eq.\,(3.14) of \cite{penguins} ---
our corrections should be implemented 
only in the calculation
of the LO QCD Wilson coefficient $C_7^{(0)eff}(\mu_b)$.

The total effect of \ew corrections on the NLO calculation of BR$_\gamma$ is a 
 $- 2.0\pm 0.3$\% reduction for a light Higgs mass, $\mh\approx 100 $ GeV.
For larger Higgs masses the effect becomes smaller: $-1.6$\% for 
$\mh=200$ GeV and $-1.3$\% for $\mh=300$ GeV.
In the {\it gaugeless} approximation and excluding the last
contributions to \equ{ew}, the net effect on BR$_\gamma$ is 
a 0.5\% reduction for $\mh\approx 100 $ GeV. 

We now calculate BR$_\gamma$ following 
 closely \cite{nlo1,noi,marciano,kagan}, updating the input
parameters and introducing only  minor refinements in the NLO analysis.
For instance, we 
evaluate the $\msbar$ top mass from the pole top mass using the
$O(\alpha_s^2)$ expression \cite{topmass}, as these corrections 
are large and their origin is distinct. Compared to the use of the 
$O(\alpha_s)$ conversion formula, this leads to a $-0.4$\% 
reduction of BR$_\gamma$. 
With respect to the detailed
analysis of \cite{noi}, we adopt a new CKM factor $f_{CKM}=|V_{ts}^*
V_{tb}/V_{cb}|^2=0.97\pm 0.02$ instead of $0.95\pm 0.03$, 
obtained using $0<\bar{\varrho}< 0.4$ from global fits of the unitarity
triangle \cite{ut}. For the semileptonic BR, we employ 
BR$_{SL}=0.1045\pm 0.0021$, corresponding to the $\Upsilon$ resonance
determination (the average of  LEP measurements is $0.1073\pm 0.0018$).
We also use $\mt\equiv\mt^{pole}=(174.3\pm5.1)$ GeV,
$\alpha_s(\mz)=0.119\pm 0.002$, $\mw= 80.419$ GeV \cite{pdg,LEP},
$\lambda_2=0.12$ GeV$^2$, $r_{cb}=m_c/m_b=0.29\pm 0.02$, $M_{cb}=
m_b-m_c=3.39\pm 0.04$ GeV.  Employing a  conventional
definition of {\it total} BR$_\gamma$ \cite{kagan} with 
a cut on the photon energy $E_\gamma> (1-\delta) \,m_b/2$,
 $\delta=0.9$, we obtain
\bea
{\rm BR}_\gamma = 0.000329 \ 
\frac{f_{CKM}}{0.97}
\frac{{\rm BR}_{SL}}{0.1045}\left(\frac{
\alpha_s(\mz)}{0.119}\right)^{1.13} 
\left(\frac{M_t}{174.3}\right)^{0.48}
\left(\frac{r_{cb}}{0.29} \right)^{0.68}
\left(\frac{M_{cb}}{3.39} \right)^{-0.3}.
\label{finalappr}
\eea
Here  the dependence of the calculation on the main input
parameters is summarized 
for small ($< 1\,\sigma$) variations around their central values.
Notice that, compared to \cite{kagan,QCD99}, the 2\% reduction due to \ew
corrections is compensated by a 2\% increase from the up-date of the
input parameters. The largest present single 
parametric uncertainty comes from $r_{cb}$ and reaches around 5\%.

The choice of $\delta=0.9$ for the photon cut--off energy in
\equ{finalappr}
is mainly motivated by the need to compare with previous literature.
The experimental measurement is based on  a much stronger cut,
$E_\gamma> 2.1$ GeV \cite{cleo}, 
but needs to be extrapolated to a more inclusive 
branching fraction (see \cite{kagan} for a recent discussion). 
On the other hand, non-perturbative problems may arise
for soft photons. 
It seems therefore useful to know BR$_\gamma$
for higher and 
more realistic cut--offs. For $0.3<\delta<0.9$ the central value of 
\equ{finalappr} is very well approximated by
\be
{\rm BR}_\gamma(\delta)=3.01 + 1.01\, \delta - 1.49 \,\delta^2  +
0.79 \,\delta^3,
\ee
which shows a mild dependence on the cut-off in a large region
of $\delta$ \cite{kagan}.

{\bf 5.} 
In summary, we have reanalyzed in detail the two-loop \ew corrections to
$B\to X_s \gamma $ and $B\to X_s g$ decays. In order to avoid dealing
with  presumably small photonic effects, in our calculation we have
neglected terms proportional to $s_\smallw^2$ not enhanced by $\mt^2$.
We have also accurately discussed the interplay between \ew and QCD
corrections. As a byproduct of the calculation we have  
presented the complete 
$O(\alpha)$ corrections to the Wilson coefficient $C_2(\mw)$.
The total effect of \ew corrections  
on the BR of $B\to X_s \gamma $, BR$_\gamma$, is  a $-2.0\pm0.3$\%
reduction, which is 
three times larger than in previous analyses \cite{strumia}. 
After inclusion of all
NLO QCD contributions, of non-pertubative corrections and of all known
QED and \ew effects, we find (for $\mh=100$ GeV and $\delta=0.9$)
\be
{\rm BR}_\gamma = (3.29\pm 0.21\pm 0.21) \times 10^{-4} 
\label{final}
\ee
where the first error is parametric and based on up-to-date
experimental inputs
and the second one is obtained by scanning the various scales
considered in \cite{noi} between $1/2$ and twice their
central values and adding 0.3\% for unaccounted \ew higher orders. 
Had we combined the scale ambiguities in quadrature, the 
second error would have been 0.16.
For $\mh=215$ GeV, at the other edge of the preferred region 
from global fits, the central value of BR$_\gamma$ is 3.30$ \times 10^{-4}$.
\equ{final} is in good agreement with the present experimental 
value BR$_\gamma=3.14\pm0.48$ \cite{cleo}.

\vspace{.6cm}
We are indebted to  M.  Misiak for pointing out to us a missing term
in \equ{ew} and for many interesting conversations
and to  A.J. Buras and  P.A. Grassi
for helpful discussions.

\end{document}